\documentclass[reprint,superscriptaddress,amsmath,amssymb,aps,prx,longbibliography,floatfix,showpacs]{revtex4-2}
\usepackage[utf8]{inputenc}
\usepackage{graphicx}
\usepackage{dcolumn}
\usepackage{bm}
\usepackage[colorlinks,citecolor=blue,linkcolor=blue,urlcolor=blue]{hyperref}
\usepackage{booktabs}
\usepackage{float}
\usepackage[T1]{fontenc}

\begin{document}

\title{Variation of Electron-electron interaction in pyrochlore structures}

\author{Jianyu Li}
\affiliation{School of Science, Harbin Institute of Technology, Shenzhen, 518055, China}
\affiliation{Shenzhen Key Laboratory of Advanced Functional Carbon Materials Research and Comprehensive Application, Shenzhen 518055, China.}

\author{Ji Liu}
\affiliation{School of Science, Harbin Institute of Technology, Shenzhen, 518055, China}
\affiliation{Shenzhen Key Laboratory of Advanced Functional Carbon Materials Research and Comprehensive Application, Shenzhen 518055, China.}

\author{Mingjun Han}
\affiliation{School of Science, Harbin Institute of Technology, Shenzhen, 518055, China}
\affiliation{Shenzhen Key Laboratory of Advanced Functional Carbon Materials Research and Comprehensive Application, Shenzhen 518055, China.}

\author{Waqas Haider}
\affiliation{School of Science, Harbin Institute of Technology, Shenzhen, 518055, China}

\author{Yusuke Nomura}
\email{yusuke.nomura@tohoku.ac.jp}
\affiliation{Institute for Materials Research, Tohoku University, Sendai, 980-8577, Japan }

\author{Ho-Kin Tang}
\email{denghaojian@hit.edu.cn}
\affiliation{School of Science, Harbin Institute of Technology, Shenzhen, 518055, China}
\affiliation{Shenzhen Key Laboratory of Advanced Functional Carbon Materials Research and Comprehensive Application, Shenzhen 518055, China.}

\date{\today}

\begin{abstract}
We conduct a comprehensive \textit{ab initio} investigation of electron-electron interactions within the pyrochlore structures of R$_2$Ru$_2$O$_7$, R$_2$Ir$_2$O$_7$, Ca$_2$Ru$_2$O$_7$, and Cd$_2$Ru$_2$O$_7$, where R denotes a rare-earth element. Utilizing a multiorbital Hubbard model, we systematically explore the effects of various rare-earth elements and applied high pressure on the correlation strength in these compounds. Our calculations on the Coulomb interaction parameter $U$ and the bandwidth $W$ reveal that the chemical pressure for R$_2$Ru$_2$O$_7$ and R$_2$Ir$_2$O$_7$ leads to an unusual increase in $U/W$ ratio, hence, increase in correlation strength.
Contrary to conventional understanding of bandwidth control, our study identifies that the Hubbard $U$ is more influential than the bandwidth $W$ behind the metal-insulator landscape of R$_2$Ru$_2$O$_7$ and R$_2$Ir$_2$O$_7$, leading to an interaction-controlled metal-insulator transition. 
We also find unexpected behavior in physical pressure. 
Whereas physical pressure leads to a decrease in the correlation strength $U/W$ as usual in R$_2$Ru$_2$O$_7$, the effect is notably small in Ca$_2$Ru$_2$O$_7$ and Cd$_2$Ru$_2$O$_7$, which provides an important clue to understanding unusual pressure-induced metal-insulator transition observed experimentally.
\end{abstract}
\pacs{Valid PACS appear here}

\maketitle

\section{Introduction}

Metal-insulator transition (MIT) has been a subject of extensive research on the strongly correlated $d$-electron systems, typically divided into three categories~\cite{RevModPhys.70.1039}: filling-controlled~\cite{ahn2016infrared,PhysRevLett.119.186803,PhysRevLett.70.2126,PhysRevLett.124.166402}, bandwidth-controlled~\cite{PhysRevLett.115.256403,PhysRevLett.90.166401,sciadv.1500797}, and dimensionality-controlled~\cite{PhysRevLett.104.147601,PhysRevLett.101.226402,PhysRevLett.124.026401} transitions.
Pyrochlore structures provide a great platform for studying 4$d$ and 5$d$ electron systems, whose electron-electron interactions remain a central focus in condensed matter physics~\cite{RevModPhys.82.53}.
These interactions lead to the emergence of various phases and transitions, including the Mott-insulating state, bad metal behavior, spin-ice-like configurations, and other forms of noncollinear magnetism, all of which can be influenced by external factors like doping~\cite{PhysRevB.102.245131,PhysRevB.103.L201111} and pressure~\cite{PhysRevB.98.075118}. 

\begin{figure}[htb!]
\begin{center}
    \includegraphics[width=\columnwidth]{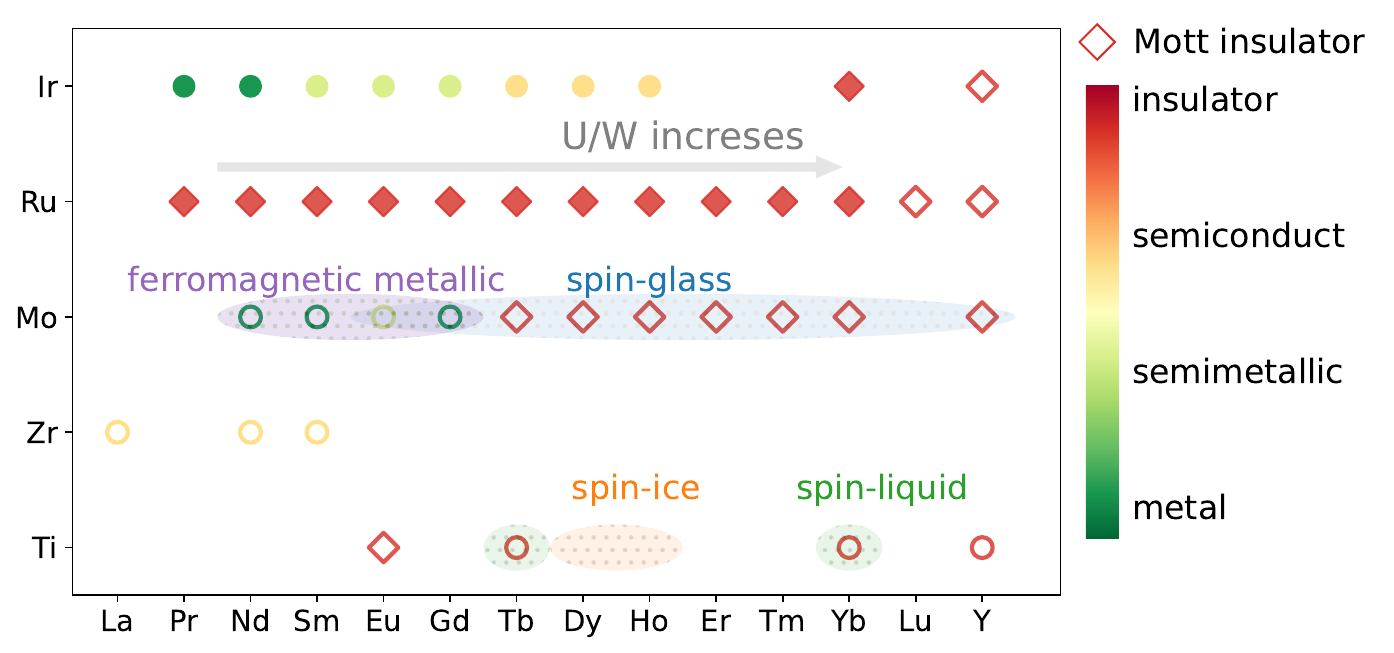}
    \caption{(Color online) A metal-insulator phase map for A$_2$B$_2$O$_7$ materials. The B-site elements include Ru, Mo, Zr, and Ti, while the A-site elements consist of various rare earth elements and Y. The map categorizes phases into insulator, semiconductor, semimetallic, and metal. Additionally, shadowed regions indicate more complex phases, such as Mott insulator, ferromagnetic metallic, spin-glass, spin-liquid, and spin-ice. Filled symbols correspond to materials for which the $U/W$ ratio was estimated and discussed in this study, whereas empty symbols denote the experimental findings without calculated ratio. Compounds that demonstrate these phase variations include
    R$_2$Ru$_2$O$_7$~\cite{Nobuyuki_Taira_1999, ITO2001337,PhysRevB.103.L201111},  R$_2$Ir$_2$O$_7$~\cite{10.1143/JPSJ.80.094701,PhysRevB.75.064426,PhysRevB.86.014428}, Y$_2$Ru$_2$O$_7$~\cite{10.1103/PhysRevB.71.060402}, Y$_2$Ir$_2$O$_7$~\cite{10.1143/JPSJ.71.2578}, R$_2$Mo$_2$O$_7$ and Y$_2$Mo$_2$O$_7$~\cite{10.1103/PhysRevLett.84.1998, 10.1103/PhysRevB.63.144425,10.1103/PhysRevB.18.2031,10.1016/0022-4596(89)90067-4}, Eu$_2$Ti$_2$O$_7$~\cite{10.1103/PhysRevB.18.2031}, Tb$_2$Ti$_2$O$_7$~\cite{10.1103/PhysRevLett.82.1012}, Y$_2$Ti$_2$O$_7$~\cite{10.1103/PhysRevB.46.5405}, Yb$_2$Ti$_2$O$_7$~\cite{10.1021/ic50017a049}.
}
\label{fig0}    
\end{center}
\end{figure}

Fig.~\ref{fig0} presents the phase map for A$_2$B$_2$O$_7$ materials, clearly delineating regions of metallic behavior, insulating states, and transition zones, unveiling the rich and complex strongly correlated phenomena inherent in these materials.
R$_2$Ru$_2$O$_7$ exhibits antiferromagnetic ordering at low temperatures and
remains electrically insulating
across all temperatures
~\cite{Nobuyuki_Taira_1999,ITO2001337}. These compounds display transitions between different states~\cite{pesin2010mott,PhysRevB.83.205101,PhysRevB.85.045124,PhysRevLett.107.127205,PhysRevLett.108.146601,PhysRevLett.109.066401,PhysRevB.82.085111}, including shifts between metallic and insulating phases, with more complex phases also being proposed~\cite{pesin2010mott,PhysRevB.82.165122}. 
In R$_2$Ir$_2$O$_7$, metallic behavior is observed in compounds with larger R$^{3+}$ ions, such as Eu, Sm, and Nd~\cite{10.1143/JPSJ.76.043706,PhysRevLett.96.087204,10.1143/JPSJ.80.094701,PhysRevB.93.245120}, while those with smaller ionic radii tend to be insulating, such as Yb and Ho~\cite{PhysRevB.86.014428,10.1143/JPSJ.80.094701,PhysRevB.93.245120}. The correlation strength in these materials can be adjusted by substituting R ions (chemical pressure) or applying physical pressure, which is a particularly intriguing aspect of pyrochlore compounds. 

Substitution of trivalent R$^{3+}$ cations with divalent cations R$^{2+}$ is also of interest. 
For instance, in the study of filling-controlled (Ca$_{1-x}$Pr$_x$)$_2$Ru$_2$O$_7$, MIT is observed~\cite{PhysRevB.103.L201111}. Pr$_2$Ru$_2$O$_7$ behaves as a Mott insulator, while Ca$_2$Ru$_2$O$_7$ is metallic. Moreover, Cd$_2$Ru$_2$O$_7$ undergoes a metal-to-insulator transition under increasing pressure~\cite{PhysRevB.98.075118,WANG19981005}, contrasting with the related monoclinic compound Hg$_2$Ru$_2$O$_7$, which transitions from a bad metal to a good metal under pressure~\cite{JPSJ.76.063707}.

These experiments under varying conditions often show unexpected behaviors, adding to the complexity of pyrochlore compounds. Here, we employ first-principles calculations to explore the underlying mechanisms driving these unusual tendencies.
The analysis of electron-electron interactions in pyrochlore structures can be broadly categorized into two regimes based on the ratio of $U/W$, where $U$ represents on-site Coulomb interaction parameter and $W$ represents the electronic bandwidth. In the weak-to-intermediate correlation regime, where $U/W \lesssim 1$, electrons remain sufficiently delocalized, allowing band topology to play a significant role in determining the material’s electronic behavior~\cite{RevModPhys.82.3045,RevModPhys.83.1057,annurev-conmatphys-062910-140432,PhysRevLett.106.236805}. 
In contrast, the strong Mott limit, where $U/W \gg 1$, is dominated by electron localization, leading to insulating behavior driven by strong correlations~\cite{PhysRevB.82.174440,PhysRevB.84.094420,PhysRevB.84.104439,10.1146/annurev-conmatphys-020911-125138}. 

We find two counterintuitive trends. First, for R$_2$Ru$_2$O$_7$ and R$_2$Ir$_2$O$_7$,  \( U \) increases as the ionic radius of R decreases, despite an arch-like behavior in the bandwidth caused by the interplay between lattice constant variations and bond angle adjustments. This highlights the dominant role of \( U \) over the bandwidth \( W \) in determining the \( U/W \) ratio, revealing a novel aspect of interaction-controlled metal-insulator transition. 
Second, in  the case of physical pressure for Ca$_2$Ru$_2$O$_7$ and Cd$_2$Ru$_2$O$_7$, the \( U/W \) ratio remains nearly constant, challenging the conventional expectation that pressure decreases \( U/W \). 

The rest of this paper is outlined as follows. 
Sec.\ \ref{Method and Model} details the derivation of electronic low-energy models using maximally localized Wannier orbitals (MLWOs) and the constrained random phase approximation (cRPA). These models, including extended multiorbital Hubbard models, are based on density functional theory (DFT) calculations, allowing precise determination of transfer integrals and effective interactions.
In Sec.\ \ref{Result1}, 
we obtained parameter results of R$_2$Ru$_2$O$_7$ and R$_2$Ir$_2$O$_7$ for the Coulomb interaction and analyzed their trends with varying ions, especially showcasing the pressure dependence of Pr$_2$Ru$_2$O$_7$ and Yb$_2$Ru$_2$O$_7$ and analyzing the differences between chemical pressure and physical pressure. 
In Sec.\ \ref{Result2}, we present the study details and obtained parameter results of Ca$_2$Ru$_2$O$_7$ and Cd$_2$Ru$_2$O$_7$ for the Coulomb interaction and analyzed their trends with varying pressure. 
In Sec.\ \ref{Conlusion}, we discuss the origin of a nontrivial behavior of $U/W$ in pyrochlore compounds and its relevance to experimental results.

\section{METHODS}
\label{Method and Model}

Here, we describe a method to derive realistic extended multi-orbital Hubbard model from first principles. We first perform band calculations based on DFT and choose ``target bands" of the effective model. 
In the case of pyrochlore compounds, we chose Ru/Ir $t_{2g}$ bands as target bands.
By constructing MLWOs for the target bands, we calculate transfer integrals and effective interactions.
In the calculation of the effective interaction, the screening by electrons besides target-band electrons is considered within the cRPA.
The resulting Hamiltonian consists of the transfer part $\mathcal{H}_t$, the Coulomb-repulsion part $\mathcal{H}_U$, and the exchange interactions and pair-hopping part $\mathcal{H}_J$ defined as
\begin{align}
\mathcal{H}=\mathcal{H}_t+\mathcal{H}_U+\mathcal{H}_J
\end{align}
where
\begin{align}
& \mathcal{H}_t=\sum_\sigma \sum_{i j} \sum_{n m} t_{n m}\left(\mathbf{R}_{i j}\right) a_{i n}^{\sigma \dagger} a_{j m}^\sigma, \label{Ht}\\
& \mathcal{H}_U=\frac{1}{2} \sum_{\sigma \rho} \sum_{i j} \sum_{n m} U_{n m}\left(\mathbf{R}_{i j}\right) a_{i n}^{\sigma \dagger} a_{j m}^{\rho \dagger} a_{j m}^\rho a_{i n}^\sigma, \\
& \mathcal{H}_J=\frac{1}{2} \sum_{\sigma \rho} \sum_{i j} \sum_{n m} J_{n m}\left(\mathbf{R}_{i j}\right) \notag\\
& \times\left(a_{i n}^{\sigma \dagger} a_{j m}^{\rho \dagger} a_{i n}^\rho a_{j m}^\sigma+a_{i n}^{\sigma \dagger} a_{i n}^{\rho \dagger} a_{j m}^\rho a_{j m}^\sigma\right)
\end{align}
with $a_{i n}^{\sigma \dagger}\left(a_{i n}^\sigma\right)$ being a creation (annihilation) operator of an electron with spin $\sigma$ in the $n$th MLWO localized at the cell located at $\mathbf{R}_i$ and $\mathbf{R}_{ij}$ = $\mathbf{R}_i - \mathbf{R}_j$.
The parameters $t_{n m}\left(\mathbf{R}_{i j}\right)$ represent on-site energy  and hopping integrals:
\begin{align}
t_{n m}(\mathbf{R})=\left\langle\phi_{n \mathbf{R}}\left|\mathcal{H}_{\mathrm{KS}}\right| \phi_{m 0}\right\rangle,
\end{align}
where $\left|\phi_{n \mathbf{R}_i}\right\rangle=a_{i n}^{\dagger}|0\rangle$ and $\mathcal{H}_{\mathrm{KS}}$ is the Kohn-Sham Hamiltonian.

We evaluate the screened Coulomb interaction $W\left(\mathbf{r}, \mathbf{r}^{\prime}\right)$ to determine the effective interaction parameters $U_{nm}(\mathbf{R})$ and $J_{nm}(\mathbf{R})$ in the low-frequency limit. This process begins by calculating the noninteracting polarization function $\chi$, excluding contributions from the target bands. It is important to account for the screening effects from the target electrons when solving the effective models, thereby avoiding double counting during the derivation. Using the calculated $\chi$, the partially screened interaction $W$ is obtained as $W = (1 - v\chi)^{-1} v$, where $v$ represents the bare Coulomb interaction, given by $v\left(\mathbf{r}, \mathbf{r}^{\prime}\right) = \frac{1}{\left|\mathbf{r}-\mathbf{r}^{\prime}\right|}$.

Once the partially screened Coulomb interaction $W\left(\mathbf{r}, \mathbf{r}^{\prime}\right)$ is calculated, the matrix elements of $W$ are obtained as
\begin{align}
U_{i j} & =\iint d \mathbf{r} d \mathbf{r}^{\prime}\left|\phi_{i \mathbf{0}}(\mathbf{r})\right|^2 W\left(\mathbf{r}, \mathbf{r}^{\prime}\right)\left|\phi_{j \mathbf{0}}\left(\mathbf{r}^{\prime}\right)\right|^2 \notag\\
& =\frac{4 \pi }{N \Omega} \sum_{\mathbf{q}} \sum_{\mathbf{G} \mathbf{G}^{\prime}} \rho_{i i}(\mathbf{q}+\mathbf{G}) W_{\mathbf{G}, \mathbf{G}^{\prime}}(\mathbf{q}) \rho_{j j}^*\left(\mathbf{q}+\mathbf{G}^{\prime}\right)
\label{U}
\end{align}
and
\begin{align}
J_{i j} & =\iint d \mathbf{r} d \mathbf{r}^{\prime} \phi_{i \mathbf{0}}^*(\mathbf{r}) \phi_{j \mathbf{0}}(\mathbf{r}) W\left(\mathbf{r}, \mathbf{r}^{\prime}\right) \phi_{j \mathbf{0}}^*\left(\mathbf{r}^{\prime}\right) \phi_{i \mathbf{0}}\left(\mathbf{r}^{\prime}\right) \notag\\
& =\frac{4 \pi }{N \Omega} \sum_{\mathbf{q}} \sum_{\mathbf{G G}^{\prime}} \rho_{i j}(\mathbf{q}+\mathbf{G}) W_{\mathbf{G}, \mathbf{G}^{\prime}}(\mathbf{q}) \rho_{i j}^*\left(\mathbf{q}+\mathbf{G}^{\prime}\right),
\label{J}
\end{align}
respectively, where $\Omega$ is the volume of the unit cell, and $\rho_{i j}(\mathbf{q}+\mathbf{G})$ is given, with the Wannier-gauge Bloch functions $\psi_{i \mathbf{k}}^{(w)}$, by
\begin{align}
    \rho_{i j}(\mathbf{q}+\mathbf{G})=\frac{1}{N} \sum_{\mathbf{k}}\left\langle\psi_{i \mathbf{k}+\mathbf{q}}^{(w)}\left|e^{i(\mathbf{q}+\mathbf{G}) \cdot \mathbf{r}}\right| \psi_{j \mathbf{k}}^{(w)}\right\rangle .
\end{align}

To facilitate a comparative analysis with the cRPA results, we compute interaction parameters using the unscreened
case, representing the Wannier matrix elements of the bare Coulomb interaction. In order to differentiate it from the cRPA results, we refer to this as the  ``bare'' interaction.

\section{results for \texorpdfstring{$\mathbf{R}_2\mathbf{Ru}_2\mathbf{O}_7$}{R2Ru2O7} and \texorpdfstring{$\mathbf{R}_2\mathbf{Ir}_2\mathbf{O}_7$}{R2Ir2O7}}
\label{Result1}

\subsection{Calculation details and lattice structures}

\begin{figure}[htb!]
\begin{center}
    \includegraphics[width=\columnwidth]{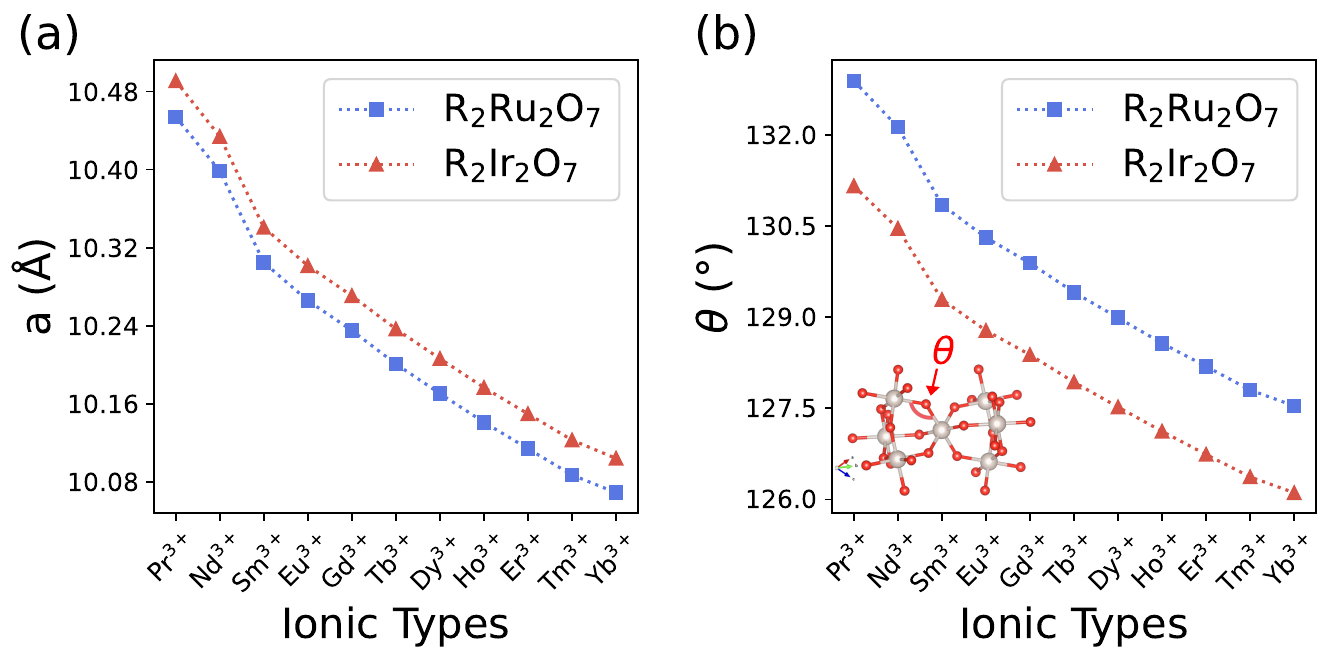}
    \caption{(Color online) Theoretically optimized values of R$_2$Ru$_2$O$_7$ and R$_2$Ir$_2$O$_7$ of (a) lattice constant and (b) bond angles of Ru-O-Ru and Ir-O-Ir. The rare-earth ions (R\(^{3+}\)) included are Pr\(^{3+}\), Nd\(^{3+}\), Sm\(^{3+}\), Eu\(^{3+}\), Gd\(^{3+}\), Tb\(^{3+}\), Dy\(^{3+}\), Ho\(^{3+}\), Er\(^{3+}\), Tm\(^{3+}\), and Yb\(^{3+}\). (Inset): The diagrammatic sketch of the bond angles. 
    }
\label{Lattice_Constant}    
\end{center}
\end{figure}

To derive the realistic extended Hubbard models for the Ru/Ir-\(\mathrm{t}_{2g}\) manifold in R$_2$Ru$_2$O$_7$/R$_2$Ir$_2$O$_7$, we utilize a combination of MLWOs \cite{wannier1,wannier2} and cRPA method \cite{cRPA}, as described in Sec.~\ref{Method and Model}. The construction of Wannier functions and the cRPA calculations are executed using the open-source package RESPACK \cite{RESPACK}, which employs a band disentanglement scheme as described in reference~\cite{PhysRevB.83.121101}. Initially, DFT band-structure calculations are performed using Quantum ESPRESSO \cite{QE}. Optimized norm-preserving Vanderbilt pseudopotentials \cite{pseudopotentials}, in conjunction with the PBE exchange–correlation functional \cite{PBE}, are sourced from PseudoDojo \cite{PseudoDojo} and employed in the DFT calculations. A $9 \times 9 \times 9$ k-point mesh is used, with an energy cutoff set to 100 Ry for the wavefunctions and 400 Ry for the electron-charge density.
The DFT calculations were conducted for the following materials: R$_2$Ru$_2$O$_7$ and R$_2$Ir$_2$O$_7$ (where R\(^{3+}\) represents Pr\(^{3+}\), Nd\(^{3+}\), Sm\(^{3+}\), Eu\(^{3+}\), Gd\(^{3+}\), Tb\(^{3+}\), Dy\(^{3+}\), Ho\(^{3+}\), Er\(^{3+}\), Tm\(^{3+}\), and Yb\(^{3+}\)). The atomic and ionic radius of lanthanide elements gradually decrease with increasing atomic number. 
For R$_2$Ru$_2$O$_7$, the Ru-\(t_{2g}\) bands are isolated from other bands, and the Wannier functions are derived from an energy range of \([-1.5, 1.5]\) eV. Using these Wannier orbitals, effective interaction parameters are computed via the cRPA method. The polarization function is calculated using 200 bands with an energy cutoff of 10 Ry.
The Brillouin-zone integral on the wave vector was evaluated by the generalized-tetrahedron method. 

Before presenting the computational results, we summarize the fundamental properties of the pyrochlore structures. 
By taking experimental crystal structure as a starting point~\cite{SUBRAMANIAN198355}, we optimize the lattice constant and internal atomic coordinates. 
Fig.~\ref{Lattice_Constant} (a) presents the theoretical lattice constants for R$_2$Ru$_2$O$_7$ and R$_2$Ir$_2$O$_7$, respectively. The theoretical and experimental lattice constants agree within 1.1\%. The lattice constant \( a \) can be controlled by chemical and external pressures. The samples are arranged in order of decreasing lattice constants. Fig.~\ref{Lattice_Constant} (b) presents the bond angles $\theta$  of Ru-O-Ru and Ir-O-Ir.  Both the lattice constant \(a\) and the bond angles $\theta$ decrease together with ion radius decreasing.

\subsection{Band structures}

\begin{figure}[htb!]
\begin{center}
    \includegraphics[width=\columnwidth]{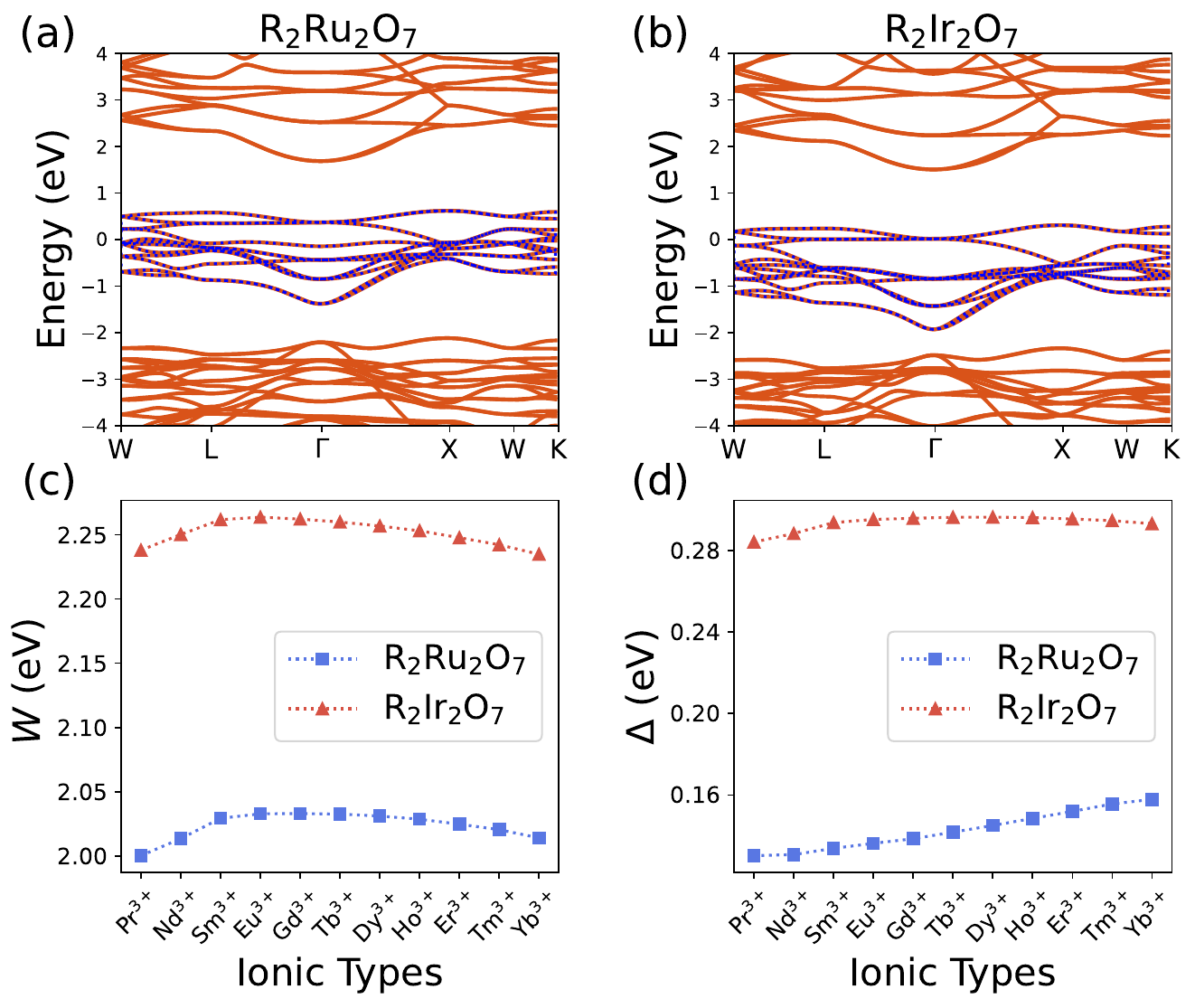}
    \caption{(Color online) Calculated $ab$ $initio$ electronic band structure of (a) R$_2$Ru$_2$O$_7$ and (b) R$_2$Ir$_2$O$_7$. The horizontal axis denotes special points in the Brillouin zone: \(W\) (0.25, 0.5, 0.75), \(L\) (0.5, 0.5, 0.5), \(\Gamma\) (0, 0, 0), \(X\) (0.5, 0, 0.5), and \(K\) (0.375, 0.375, 0.75). The interpolated band dispersion, derived from the Wannier tight-binding Hamiltonian, is depicted by blue dashed lines. The rare-earth ions dependence for R$_2$Ru$_2$O$_7$ and R$_2$Ir$_2$O$_7$ of (c) the bandwidth \(W\) of the Ru-\(t_{2g}\) band and (d) the crystal field splitting \(\Delta\) between the $e_g^{\prime}$ and \(a_{1g}\) orbitals within the \(t_{2g}\) manifold.}
\label{bands}    
\end{center}
\end{figure}

The calculated bands, depicted as red lines in Fig.~\ref{bands}(a-b), are based on structures for R$_2$Ru$_2$O$_7$ and R$_2$Ir$_2$O$_7$~\cite{10.1143/JPSJ.75.103801, YAMAMOTO1994372}. R$_2$Ru$_2$O$_7$ and R$_2$Ir$_2$O$_7$ exhibit highly similar band structures. MLWOs~\cite{wannier1,wannier2} are constructed by projecting the Ru/Ir-\(t_{2g}\) orbitals. The resultant bands, illustrated as dashed blue lines in Fig.~\ref{bands}(a-b), are aligned around the Fermi energy, set at 0 eV.
The dominant crystal field splitting comes from the oxygen octahedra surrounding each Ru$^{4+}$/Ir$^{4+}$ cation, which splits the levels into a higher-energy $e_g$ orbital doublet and a lower $t_{2 g}$ orbital triplet, spanned by orbitals with $x y$, $y z$, and $z x$ symmetry. These are separated by an $\sim 2 \mathrm{eV}$ gap, and as such we can neglect the higher-energy $e_g$ levels. 

Fig.~\ref{bands}(c) shows nontrivial arch-like bandwidth behavior.
In R$_2$Ru$_2$O$_7$ and R$_2$Ir$_2$O$_7$, this variation in R has been identified as a key factor influencing the bandwidth. 
Intuitively, one might expect a smaller R$^{3+}$ radius, and hence a smaller lattice constant, to result in a larger bandwidth. 
However, as the ionic radius of R$^{3+}$ decreases, the decline of bond angles contributes the bandwidth $W$ decreasing [Fig.~\ref{Lattice_Constant}(b)].
Consequently,  simultaneous changes in the lattice constant and Ru-O-Ru and Ir-O-Ir bond angles counteract each effect. We find that due to a compensating effect between the lattice constant change and the bond angle adjustments, the bandwidth exhibits an arch-like behavior and remains almost unchanged overall.
This bandwidth behavior can also be preliminarily explained by noting that larger R$^{3+}$ ions reduce the trigonal compression of the octahedra, increase the overlap of Ru-O and Ir-O orbitals, and thereby promote the transition of Ru and Ir electrons~\cite{KOO1998269}.

\subsection{MLWOs and Transfer Integrals}
\begin{figure}[htb!]
\begin{center}
    \includegraphics[width=0.75\columnwidth]{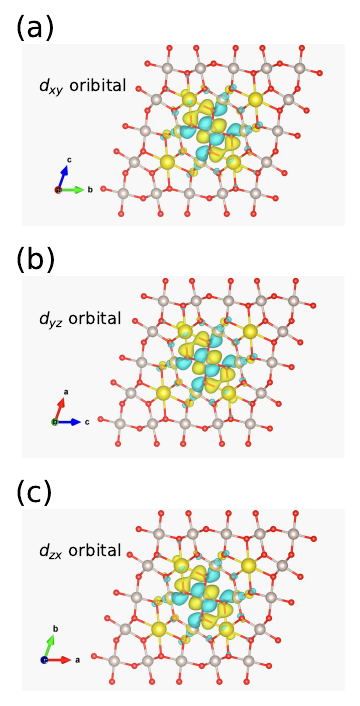}
    \caption{(Color online) Isosurfaces of MLWOs of of R$_2$Ru$_2$O$_7$ including (a) \(d_{xy}\), (b) \(d_{yz}\) and (c) \(d_{zx}\) orbitals, drawn by VESTA~\cite{Momma:ko5060}. The yellow and blue surfaces indicate positive and negative isosurfaces respectively, and the red and grey balls indicate O atom and Ru atom respectively. Note that the three panels are depicted using different crystal orientations.}
\label{wannier}    
\end{center}
\end{figure}

\newlength{\oldtabcolsep}
\setlength{\oldtabcolsep}{\tabcolsep}
\setlength{\tabcolsep}{8pt}
\begin{table}[!htb]
\renewcommand{\arraystretch}{1}
    \caption{Hopping parameters for R$_2$Ru$_2$O$_7$ in Eq.~(\ref{matrix}).  Units are given in eV.}
    \centering
    \fontsize{8}{12}\selectfont
    \resizebox{0.5\textwidth}{!}{%
    \begin{tabular}{clllll}
    \hline\hline
    \multicolumn{1}{c}{Parameters} & \multicolumn{1}{c}{F$_1$} & \multicolumn{1}{c}{F$_2$} & \multicolumn{1}{c}{F$_3$} & \multicolumn{1}{c}{F$_4$} \\
    \hline
    Pr$_2$Ru$_2$O$_7$ & 0.0151 & $-$0.1025 & 0.0059 & 0.1543 \\
    Nd$_2$Ru$_2$O$_7$ & 0.0157 & $-$0.1059 & 0.0034 & 0.1501 \\
    Sm$_2$Ru$_2$O$_7$ & 0.0165 & $-$0.1117 & $-$0.0012 & 0.1418 \\
    Eu$_2$Ru$_2$O$_7$ & 0.0167 & $-$0.1140 & $-$0.0032 & 0.1376 \\
    Tb$_2$Ru$_2$O$_7$ & 0.0170 & $-$0.1177 & $-$0.0068 & 0.1296 \\
    Dy$_2$Ru$_2$O$_7$ & 0.0170 & $-$0.1195 & $-$0.0085 & 0.1255 \\
    Ho$_2$Ru$_2$O$_7$ & 0.0171 & $-$0.1211 & $-$0.0102 & 0.1212 \\
    Er$_2$Ru$_2$O$_7$ & 0.0171 & $-$0.1225 & $-$0.0119 & 0.1170 \\
    Tm$_2$Ru$_2$O$_7$ & 0.0172 & $-$0.1240 & $-$0.0135 & 0.1127 \\
    Yb$_2$Ru$_2$O$_7$ & 0.0170 & $-$0.1247 & $-$0.0145 & 0.1091 \\
    \hline\hline   
    \end{tabular}%
    }
\label{table3}
\end{table}

\begin{table}[!htb]
\renewcommand{\arraystretch}{1}
    \caption{Hopping parameters for R$_2$Ir$_2$O$_7$ in Eq.~(\ref{matrix}). Units are given in eV.}
    \centering
    \fontsize{8}{12}\selectfont
    \resizebox{0.5\textwidth}{!}{%
    \begin{tabular}{clllll}
    \hline\hline
    \multicolumn{1}{c}{Parameters} & \multicolumn{1}{c}{F$_1$} & \multicolumn{1}{c}{F$_2$} & \multicolumn{1}{c}{F$_3$} & \multicolumn{1}{c}{F$_4$} \\
    \hline
    Pr$_2$Ir$_2$O$_7$ & 0.0099 & $-0.1148$ & 0.0007 & 0.1631 \\
    Nd$_2$Ir$_2$O$_7$ & 0.0104 & $-$0.1176 & $-$0.0019 & 0.1595 \\
    Sm$_2$Ir$_2$O$_7$ & 0.0112 & $-$0.1224 & $-$0.0063 & 0.1515 \\
    Eu$_2$Ir$_2$O$_7$ & 0.0114 & $-$0.1244 & $-$0.0083 & 0.1473 \\
    Tb$_2$Ir$_2$O$_7$ & 0.0118 & $-$0.1275 & $-$0.0117 & 0.1390 \\
    Dy$_2$Ir$_2$O$_7$ & 0.0119 & $-$0.1290 & $-$0.0133 & 0.1346 \\
    Ho$_2$Ir$_2$O$_7$ & 0.0120 & $-$0.1304 & $-$0.0149 & 0.1301 \\
    Er$_2$Ir$_2$O$_7$ & 0.0121 & $-$0.1317 & $-$0.0163 & 0.1255 \\
    Tm$_2$Ir$_2$O$_7$ & 0.0122 & $-$0.1329 & $-$0.0178 & 0.1208 \\
    Yb$_2$Ir$_2$O$_7$ & 0.0121 & $-$0.1336 & $-$0.0188 & 0.1169 \\
    \hline\hline   
    \end{tabular}%
    }
\label{table4}
\end{table}
\setlength{\tabcolsep}{\oldtabcolsep}

We constructed 12 Wannier orbitals corresponding to the Ru/Ir \(t_{2g}\) [Fig.~\ref{wannier}]. This set includes 3 \(t_{2g}\) orbitals for each of the 4 Ru/Ir atoms, serving as the basis for the Kohn-Sham Hamiltonian matrix elements. The transfer integrals \(t_{nm}(\mathbf{R})\) are represented as \(12 \times 12\) matrices. Each \(t_{2g}\) orbital, indexed as 1, 2, and 3, corresponds to \(d_{xy}\)-, \(d_{yz}\)-, and \(d_{zx}\)-like orbitals, respectively. 

Fig.~\ref{wannier} presents a contour plot of one of the MLWOs associated with the $t_{2g}$ bands of R$_2$Ru$_2$O$_7$. The results for R$_2$Ir$_2$O$_7$ exhibit a high degree of similarity.
From the figure, it is evident that the resulting Wannier orbital shows three distinctive cloverleaf shapes within the $xy$-, $yz$- and $zx$-plane which represent $d_{xy}$-like, $d_{yz}$-like and $d_{zx}$-like symmetry.
We also note that the $t_{2g}$ orbitals of a Ru/Ir atom overlap with surrounding O-$2p$ orbitals.

Transfer integrals to a particular site can be mapped to equivalent sites using appropriate symmetry operations.
In the \(\mathbf{R} = (R_x, R_y, R_z) = (0, 0, 0)\), there are four Ru or Ir atoms. Given the fcc structure, we consider the hopping between the first Ru or Ir atom located at $(0, 0, 0)$ and the second one located at $(0, 1/4, 1/4)$ as the nearest-neighbor site hopping, where the coordinate is based on the conventional cell. The transfer matrices to these sites are

\begin{align}
\left( \begin{matrix}
	F_1 & F_3 & F_3 \\
	F_3 & F_2 & F_4 \\
	F_3 & F_4 & F_2 \\
\end{matrix} \right) .
\label{matrix}
\end{align}

Tables \ref{table3} and \ref{table4} present the values of $F_1$ to $F_4$ for R$_2$Ru$_2$O$_7$ and R$_2$Ir$_2$O$_7$, respectively. 
For R$_2$Ru$_2$O$_7$ and R$_2$Ir$_2$O$_7$, as the ionic radius of R decreases (from Pr to Yb), the absolute value of \(F_1\) and \(F_2\) demonstrate a gradual increase in magnitude, while \(F_3\) and \(F_4\) show a decline trend, including sign changes for \(F_3\).

Using these transfer parameters, we constructed the transfer term \( H_t \) in Eq.~\ref{Ht} of the effective model. The band dispersion for the pyrochlore compounds, calculated from this \( H_t \), is shown as blue dashed lines in Fig.~\ref{bands}(a-b). This demonstrates that the original band dispersion is satisfactorily reproduced.

\begin{table*}[!htb]
\renewcommand{\arraystretch}{1}
    \caption{$U$, $U^{\prime}$ and $J$ with different screening levels [unscreened (bare) and constrained RPA (cRPA)] for R$_2$Ru$_2$O$_7$, where R\(^{3+}\) represents Pr\(^{3+}\), Nd\(^{3+}\), Sm\(^{3+}\), Eu\(^{3+}\), Gd\(^{3+}\), Tb\(^{3+}\), Dy\(^{3+}\), Ho\(^{3+}\), Er\(^{3+}\), Tm\(^{3+}\), and Yb\(^{3+}\). Units are given in eV. At the bottom of the table, we present our calculated cRPA-macroscopic-dielectric constant $\epsilon _{M}^{\mathrm{cRPA}}$ in Eq.~(\ref{epsilon}).}
    \centering
    \resizebox{1\textwidth}{1.75cm}{%
    \begin{tabular}{cllllllllllllllllllllllllllll}
    \hline\hline
     \multicolumn{1}{c}{materials} & \multicolumn{1}{c}{Pr$_2$Ru$_2$O$_7$} & \multicolumn{1}{c}{Nd$_2$Ru$_2$O$_7$} & \multicolumn{1}{c}{Sm$_2$Ru$_2$O$_7$} & \multicolumn{1}{c}{Eu$_2$Ru$_2$O$_7$} & \multicolumn{1}{c}{Gd$_2$Ru$_2$O$_7$} & \multicolumn{1}{c}{Tb$_2$Ru$_2$O$_7$} & \multicolumn{1}{c}{Dy$_2$Ru$_2$O$_7$} & \multicolumn{1}{c}{Ho$_2$Ru$_2$O$_7$} & \multicolumn{1}{c}{Er$_2$Ru$_2$O$_7$} & \multicolumn{1}{c}{Tm$_2$Ru$_2$O$_7$} & \multicolumn{1}{c}{Yb$_2$Ru$_2$O$_7$} \\
        \hline
    $U_{\mathrm{bare}}$	&	11.492 	&	11.549 	&	11.636 	&	11.670 	&	11.695 	&	11.721 	&	11.742 	&	11.760 	&	11.774 	&	11.786 	&	11.794 	\\
    $U^{\prime}_{\mathrm{bare}}$	&	10.545 	&	10.598 	&	10.679 	&	10.709 	&	10.733 	&	10.757 	&	10.777 	&	10.793 	&	10.806 	&	10.818 	&	10.825 	\\
    $U_{\mathrm{cRPA}}$	&	2.393 	&	2.426 	&	2.480 	&	2.502 	&	2.519 	&	2.538 	&	2.553 	&	2.566 	&	2.578 	&	2.590 	&	2.599 	\\
    $U^{\prime}_{\mathrm{cRPA}}$	&	1.637 	&	1.664 	&	1.710 	&	1.728 	&	1.743 	&	1.759 	&	1.772 	&	1.782 	&	1.793 	&	1.803 	&	1.811 	\\
    $J_{\mathrm{bare}}$	&	0.410 	&	0.413 	&	0.418 	&	0.420 	&	0.421 	&	0.423 	&	0.424 	&	0.424 	&	0.425 	&	0.426 	&	0.426 	\\
    $J_{\mathrm{cRPA}}$	&	0.341 	&	0.344 	&	0.349 	&	0.351 	&	0.352 	&	0.353 	&	0.355 	&	0.355 	&	0.356 	&	0.357 	&	0.357 	\\
    $\epsilon _{M}^{\mathrm{cRPA}}$	&	7.23 	&	7.15 	&	7.03 	&	6.99 	&	6.95 	&	6.91 	&	6.89 	&	6.88 	&	6.86 	&	6.85 	&	6.83 	\\
    \hline\hline   
        \end{tabular}%
    }
\label{table1}
\end{table*}

\begin{table*}[!htb]
\renewcommand{\arraystretch}{1}
    \caption{$U$, $U^{\prime}$ and $J$ with different screening levels [unscreened (bare) and constrained RPA (cRPA)] for R$_2$Ir$_2$O$_7$, where R\(^{3+}\) represents Pr\(^{3+}\), Nd\(^{3+}\), Sm\(^{3+}\), Eu\(^{3+}\), Gd\(^{3+}\), Tb\(^{3+}\), Dy\(^{3+}\), Ho\(^{3+}\), Er\(^{3+}\), Tm\(^{3+}\), and Yb\(^{3+}\). Units are given in eV. At the bottom of the table, we present our calculated cRPA-macroscopic-dielectric constant $\epsilon _{M}^{\mathrm{cRPA}}$ in Eq.~(\ref{epsilon}).}
    \centering
    \resizebox{1\textwidth}{1.75cm}{%
    \begin{tabular}{cllllllllllllllllllllllllllll}
    \hline\hline
     \multicolumn{1}{c}{materials} & \multicolumn{1}{c}{Pr$_2$Ir$_2$O$_7$} & \multicolumn{1}{c}{Nd$_2$Ir$_2$O$_7$} & \multicolumn{1}{c}{Sm$_2$Ir$_2$O$_7$} & \multicolumn{1}{c}{Eu$_2$Ir$_2$O$_7$} & \multicolumn{1}{c}{Gd$_2$Ir$_2$O$_7$} & \multicolumn{1}{c}{Tb$_2$Ir$_2$O$_7$} & \multicolumn{1}{c}{Dy$_2$Ir$_2$O$_7$} & \multicolumn{1}{c}{Ho$_2$Ir$_2$O$_7$} & \multicolumn{1}{c}{Er$_2$Ir$_2$O$_7$} & \multicolumn{1}{c}{Tm$_2$Ir$_2$O$_7$} & \multicolumn{1}{c}{Yb$_2$Ir$_2$O$_7$} \\
        \hline
    $U_{\mathrm{bare}}$	&	10.335	&	10.396	&	10.493	&	10.531	&	10.561	&	10.591	&	10.615	&	10.636	&	10.654	&	10.669	&	10.679	\\
    $U^{\prime}_{\mathrm{bare}}$	&	9.451	&	9.509	&	9.598	&	9.633	&	9.661	&	9.689	&	9.711	&	9.731	&	9.747	&	9.761	&	9.77	\\
    $U_{\mathrm{cRPA}}$	&	2.317	&	2.357	&	2.421	&	2.447	&	2.468	&	2.49	&	2.509	&	2.526	&	2.541	&	2.554	&	2.564	\\
    $U^{\prime}_{\mathrm{cRPA}}$	&	1.637	&	1.671	&	1.725	&	1.747	&	1.764	&	1.783	&	1.799	&	1.813	&	1.826	&	1.837	&	1.845	\\
    $J_{\mathrm{bare}}$	&	0.361	&	0.364	&	0.37	&	0.372	&	0.373	&	0.375	&	0.376	&	0.377	&	0.378	&	0.379	&	0.379	\\
    $J_{\mathrm{cRPA}}$	&	0.29	&	0.294	&	0.299	&	0.301	&	0.303	&	0.304	&	0.306	&	0.307	&	0.308	&	0.309	&	0.309	\\
    $\epsilon _{M}^{\mathrm{cRPA}}$	&	6.91 	&	6.84 	&	6.71 	&	6.66 	&	6.62 	&	6.58 	&	6.54 	&	6.51 	&	6.48 	&	6.45 	&	6.44 	\\
    \hline\hline   
        \end{tabular}%
    }
\label{table2}
\end{table*}

\subsection{Effective Interaction Parameters}

\begin{figure}
\begin{center}
    \includegraphics[width=\columnwidth]{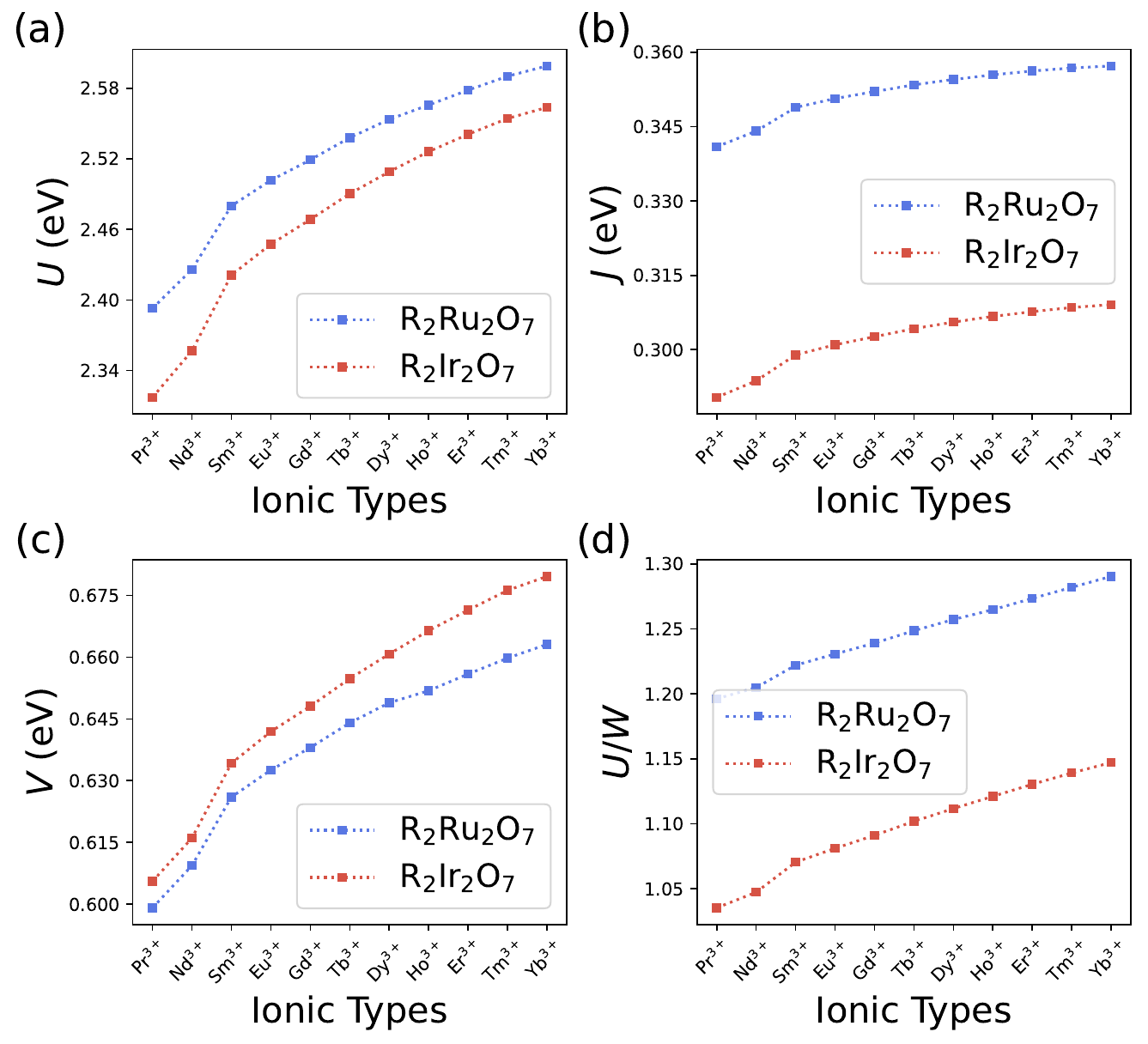}
    \caption{(Color online) The dependence of various electronic parameters on trivalent rare-earth ions for R$_2$Ru$_2$O$_7$ and R$_2$Ir$_2$O$_7$: (a) the on-site effective Coulomb repulsion \(U\), (b) the on-site effective exchange interaction $J$, (c) the off-site effective Coulomb repulsion between neighboring sites $V$, and (d) the correlation strength $U / W$, which are derived within the cRPA method.
    The data points in each panel correspond to the following rare-earth ions: Pr\(^{3+}\), Nd\(^{3+}\), Sm\(^{3+}\), Eu\(^{3+}\), Gd\(^{3+}\), Tb\(^{3+}\), Dy\(^{3+}\), Ho\(^{3+}\), Er\(^{3+}\), Tm\(^{3+}\), and Yb\(^{3+}\).
}
\label{R2B2O72}    
\end{center}
\end{figure}

Then we analyze the effective Coulomb interaction parameters for R$_2$Ru$_2$O$_7$ and R$_2$Ir$_2$O$_7$. 
$U$ is the intraorbital Coulomb interaction, $U^{\prime}$ is the interorbital interaction, and $J$ is the exchange interactions.
Table.~\ref{table1} and Table.~\ref{table2} show our calculated interaction parameters $U$, $U^{\prime}$ and $J$ with two screening levels (bare and cRPA) for R$_2$Ru$_2$O$_7$ and R$_2$Ir$_2$O$_7$ respectively.
It can be observed that the values of effective interaction parameters decrease due to the screening process. In the R$_2$Ru$_2$O$_7$ compounds, the bare Coulomb repulsion is around 11.6 eV, but considering the cRPA screening effects, this value decreases to around 2.5 eV. 
It should be noted here that the material dependence of bare values in Ru$^{4+}$ and Ir$^{4+}$ are big, for example, 11.5 eV for Pr$_2$Ru$_2$O$_7$ and 10.3 eV for Pr$_2$Ir$_2$O$_7$. The difference is beyond 11\%. On the other hand, the difference in the cRPA values is around 3\%: 2.4 eV for Pr$_2$Ru$_2$O$_7$ and 2.3 eV for Pr$_2$Ir$_2$O$_7$.
Notably, the bare values for the fcc pyrochlore structures exhibit minimal rare earth ion dependence, with values of 11.5 eV for Pr$_2$Ru$_2$O$_7$ and 11.8 eV for Yb$_2$Ru$_2$O$_7$, a difference of nearly 2.5\%. In contrast, the cRPA values show the dependence exceeding 8\%, with 2.4 eV for Pr$_2$Ru$_2$O$_7$ and 2.6 eV for Yb$_2$Ru$_2$O$_7$.
Indeed, this difference originates from the difference in screening, whose strength can be inferred from the macroscopic dielectric constant defined as
\begin{align}
\epsilon _{M}^{\mathrm{cRPA}} = \lim_{\mathbf{Q} \rightarrow 0, \omega \rightarrow 0} \frac{1}{\epsilon _{\mathbf{GG}}^{\mathrm{cRPA}^{-1}}(\mathbf{q}, \omega)},
\label{epsilon}
\end{align}
where $\omega$ is the frequency, $\mathbf{Q} = \mathbf{q} + \mathbf{G}$, with $\mathbf{q}$ being the wave vector in the first Brillouin zone, and $\mathbf{G}$ being the reciprocal lattice vector. 
We list the values at the bottom of tables. We see that the material dependence of  $\epsilon _{M}^{\mathrm{cRPA}}$ is appreciable as 7.2 for Pr$_2$Ru$_2$O$_7$ and 6.9 for Pr$_2$Ir$_2$O$_7$ respectively, clearly indicating the importance of the screening effect.

Fig.~\ref{R2B2O72} presents the results of the cRPA calculations, highlighting key parameters such as the on-site Coulomb repulsion $U$, the off-site interaction $V$ averaged over the nearest-neighbor sites, the on-site exchange interaction $J$, and the ratio $U/W$, which indicates the correlation strength within the system.
A decrease in the ionic radius of R$^{3+}$ leads to a reduction in the lattice constant of R$_2$Ru$_2$O$_7$ and R$_2$Ir$_2$O$_7$ [Fig.~\ref{Lattice_Constant}(a)]. 
Interestingly $U$ increases and this behavior can be explained by corresponding increase in bare $U$ and decrease in the dielectric constant (see Table~\ref{table1} and Table~\ref{table2}). As the dielectric constant declines, the screening effect is weakened, resulting in electrons experiencing stronger direct interactions with one another. This leads to an increase in on-site Coulomb repulsion, where electrons occupying the same or nearby atomic sites encounter a more pronounced repulsive force. Kaneko et al.~\cite{PhysRevB.102.041114} demonstrated experimentally that the magnitude of the charge gap of R$_2$Ru$_2$O$_7$ measured at 10 K increases and the peak of the Hubbard band shifts upwards as R ionic radius decreases from Pr to Lu. 
These observations align with the increasing trend of theoretical $U$.

The metal-insulator transition in correlated systems is often analyzed through the $U/W$ ratio~\cite{RevModPhys.68.13}. In a system with approximately one electron per lattice site, a Mott insulating state emerges when the Coulomb repulsion $U$ exceeds the one-particle bandwidth $W$. Conversely, a metallic state is expected when $U/W \ll 1$. When $U/W \gg 1$, the significant energy cost associated with double occupancy at a site restricts the mobility of electrons, leading to electron localization. Thus, a first-order metal-insulator transition occurs at a critical $U/W$ ratio close to 1.

Fig.~\ref{R2B2O72}(d) illustrates the variation of the $U/W$ ratio across different rare-earth ions. As the ionic radius of the rare-earth element decreases, the $U/W$ value increases. This trend is driven by the increase in on-site Coulomb interaction $U$, which outweighs the corresponding change in the bandwidth $W$. 
The primary role of $U$ in determining $U/W$ is of great interest since usually we expect $W$ plays a crucial role in the material dependence. 
However, in the case of R$_2$Ru$_2$O$_7$ and R$_2$Ir$_2$O$_7$, the parameter $U$ primarily determines whether the material exhibits metallic or insulating behavior.
Specifically, in R$_2$Ir$_2$O$_7$, the $U/W$ ratio increases as the ionic radius of the rare-earth element decreases, consistent with the observed insulating behavior in compounds containing smaller R$^{3+}$ ions, such as Yb and Ho~\cite{PhysRevB.86.014428,10.1143/JPSJ.80.094701}. 
R$_2$Ir$_2$O$_7$ is metallic for large R$^{3+}$ ionic radii with a low $U/W$ ratio ($U/W < 1.1$)~\cite{PhysRevLett.96.087204,RevModPhys.68.13}.
In contrast, R$_2$Ru$_2$O$_7$ consistently exhibits a higher $U/W$ ratio, resulting in stronger electronic correlations and insulating behavior. For instance, Pr$_2$Ru$_2$O$_7$, characterized by a high $U/W$ ratio ($U/W > 1.1$), exhibits Mott insulating behavior~\cite{PhysRevB.103.L201111}.  
The analysis above highlights that in these compounds, the $U$ parameter has a greater influence on the $U/W$ ratio than $W$, which challenges the conventional belief that changes in $W$ are of primary importance.

\subsection{Pressure dependence for \texorpdfstring{$\mathbf{Pr}_2\mathbf{Ru}_2\mathbf{O}_7$}{Pr2Ru2O7} and \texorpdfstring{$\mathbf{Yb}_2\mathbf{Ru}_2\mathbf{O}_7$}{Yb2Ru2O7}}

\begin{figure}[b!]
\includegraphics[width=1\columnwidth]{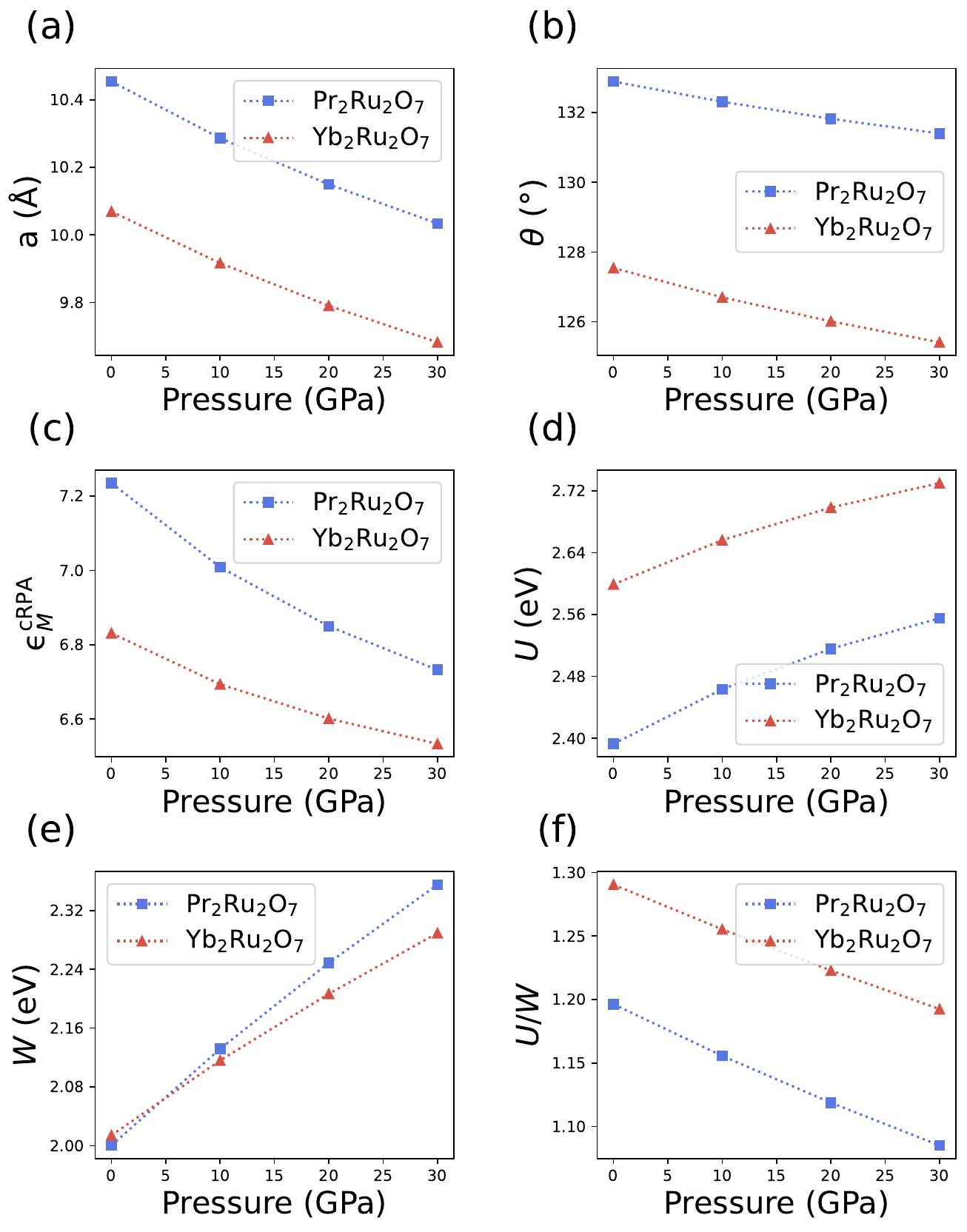}
    \caption{(Color online) The pressure dependence of various physical properties of Pr$_2$Ru$_2$O$_7$ and Yb$_2$Ru$_2$O$_7$. The properties analyzed include (a) the lattice parameter $a$, (b) bond angle $\theta$, (c) cRPA-macroscopic-dielectric constant $\epsilon _{M}^{\mathrm{cRPA}}$, (d) on-site Coulomb interaction $U$, (e) bandwidth $W$, and (f) the ratio of the correlation strength $U/W$. The $x$-axis in each plot represents the applied pressure in GPa.}
\label{plus_press}
\end{figure}

Fig.~\ref{plus_press} offers a detailed examination of how pressure affects both the structural and electronic properties of Pr$_2$Ru$_2$O$_7$ and Yb$_2$Ru$_2$O$_7$. As anticipated, the Ru-O-Ru bond angles show minimal variation with increasing pressure, while the lattice constant \(a\) decreases and the bandwidth \(W\) expands. Notably, there is a significant increase in the on-site Coulomb interaction \(U\) as pressure increases.
This rise in \(U\) under pressure can be attributed to the same mechanism responsible for the increase in \(U\) with decreasing ion radius, as discussed earlier.
Both phenomena are driven by a reduction in the macroscopic dielectric constant, which enhances electron-electron repulsion by reducing screening effects, as well as an increase in the bare $U$ value. 
In the case of physical pressure, the bandwidth $W$ monotonically increases, in contrast with chemical pressure. 
This difference arises because physical pressure causes a smaller change in Ru-O-Ru bond angles, reducing the counteracting effects. For a similar change in lattice constant, the reduction in bond angles due to physical pressure is much smaller than that caused by chemical pressure. As a result, the decrease in bandwidth from bond angle reduction is insufficient to offset the increase in bandwidth from lattice constant compression.
Since the increase in $W$ outpaces that of \(U\), resulting in a decrease in the ratio \(U/W\). Experimentally, Pr$_2$Ru$_2$O$_7$ remains an insulator under applied pressures below 20 GPa~\cite{PhysRevB.102.041114}. However, from a theoretical standpoint, the reduction in \(U/W\) at higher pressures ($>$ 20 GPa) suggests a trend towards more metallic behavior.

\section{results for \texorpdfstring{$\mathbf{Ca}_2\mathbf{Ru}_2\mathbf{O}_7$}{Ca2Ru2O7} and \texorpdfstring{$\mathbf{Cd}_2\mathbf{Ru}_2\mathbf{O}_7$}{Cd2Ru2O7}}
\label{Result2}

\subsection{Calculation details and the band structures}

\begin{figure}[htb!]
\begin{center}
    \includegraphics[width=\columnwidth]{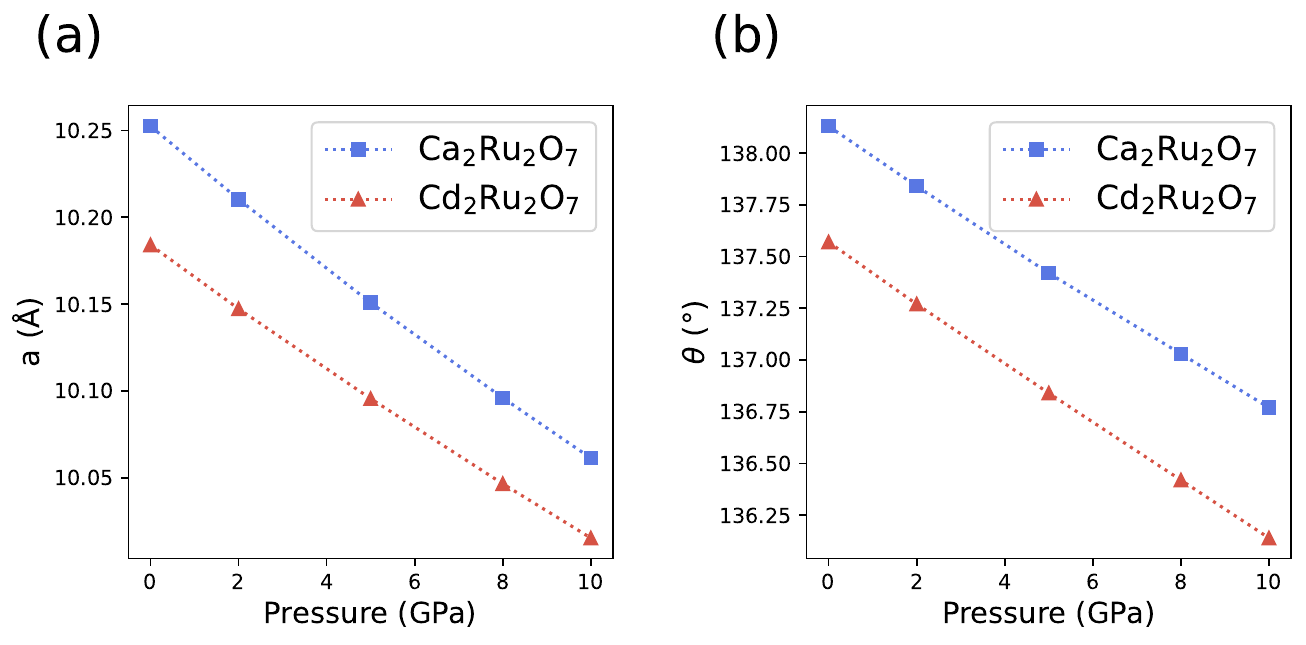}
    \caption{(Color online) The pressure dependence of basic properties for Ca$_2$Ru$_2$O$_7$ and Cd$_2$Ru$_2$O$_7$: (a) the lattice constant \(a\), (b) the bond angles $\theta$  of Ru-O-Ru.  
    }
\label{fig4}    
\end{center}
\end{figure}

\begin{figure}[htb!]
\begin{center}
    \includegraphics[width=\columnwidth]{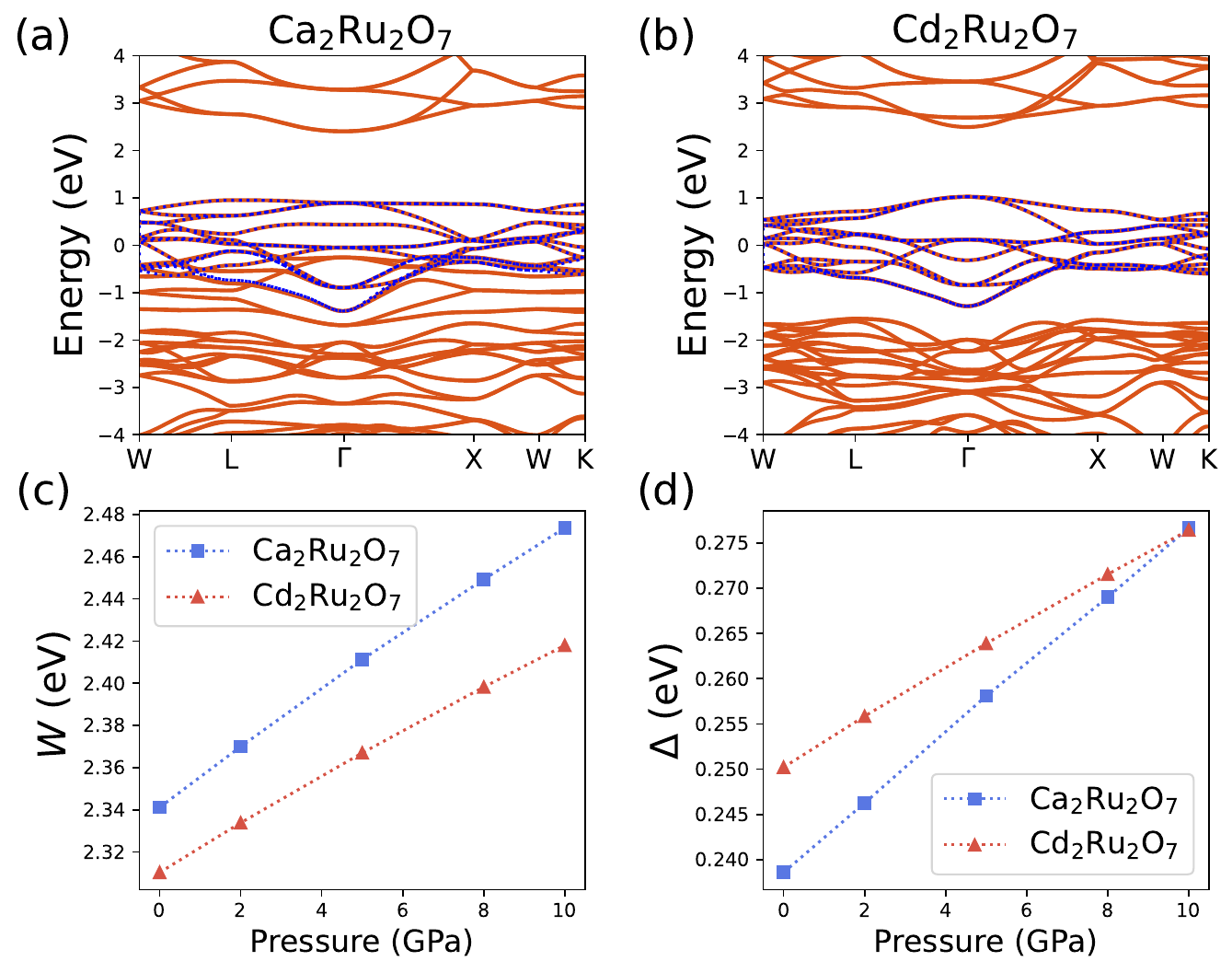}
    \caption{(Color online) Calculated $ab$ $initio$ electronic band structure of (a) Ca$_2$Ru$_2$O$_7$ and (b) Cd$_2$Ru$_2$O$_7$. The horizontal axis denotes special points in the Brillouin zone: \(W\) (0.25, 0.5, 0.75), \(L\) (0.5, 0.5, 0.5), \(\Gamma\) (0, 0, 0), \(X\) (0.5, 0, 0.5), and \(K\) (0.375, 0.375, 0.75). The interpolated band dispersion, derived from the tight-binding Hamiltonian, is depicted by blue dashed lines. The pressure dependence for Ca$_2$Ru$_2$O$_7$ and Cd$_2$Ru$_2$O$_7$ of (c) the bandwidth \(W\) of the Ru-\(t_{2\mathrm{g}}\) band and (d) the crystal field splitting \(\Delta\) between the \(e_g^{\prime}\) and \(a_{1g}\) orbitals within the \(t_{2g}\) manifold.}
\label{fig5}    
\end{center}
\end{figure}

The majority of the calculation conditions for Ca\(_2\)Ru\(_2\)O\(_7\) and Cd\(_2\)Ru\(_2\)O\(_7\) in DFT are identical to those used for R\(_2\)Ru\(_2\)O\(_7\). 
From the experimental crystal structure for Ca$_2$Ru$_2$O$_7$ and Cd$_2$Ru$_2$O$_7$~\cite{10.1143/JPSJ.75.103801,PhysRevB.98.075118}, we optimize lattice constants and internal coordinates.
These calculated values closely match the experimental values for Cd$_2$Ru$_2$O$_7$ at various pressures, with a difference of less than 1\%~\cite{PhysRevB.98.075118}.
Fig.~\ref{fig4} (a) presents the theoretical lattice constants for Ca$_2$Ru$_2$O$_7$ and Cd$_2$Ru$_2$O$_7$, respectively.  The lattice constant \( a \) can be controlled by chemical and external pressures.  Fig.~\ref{fig4} (b) presents the bond angles $\theta$  of Ru-O-Ru.  Both the lattice constant \(a\) and the bond angles $\theta$ decrease together with pressure increasing.

The Ru-\(t_{2g}\) bands of \(\mathrm{Ca}_2 \mathrm{Ru}_2 \mathrm{O}_7\) overlap with the O-$2p$ bands as depicted in Fig.~\ref{fig5}(a). Therefore, we employ both outer and inner windows with energy ranges of \([-1.45, 1.5]\) eV and \([-0.18, 1.5]\) eV, respectively.

The calculated bands, depicted as red lines in Fig.~\ref{fig5}(a-b), are based on structures for Ca$_2$Ru$_2$O$_7$ and Cd$_2$Ru$_2$O$_7$.
The parameter \(W\) typically increases with applied pressure due to the reduction in interatomic distances, which enhances orbital overlap and broadens electronic bands. Fig.~\ref{fig5}(c) illustrates this increase in \(W\), which generally leads to a transition toward more weekly correlated behavior under pressure. Conversely, Fig.~\ref{bands}(d) shows the pressure-induced increase in trigonal distortion \(\Delta\), which is associated with a transition toward more insulating behavior.

\subsection{Comparison between \texorpdfstring{$\mathbf{Ca}_2\mathbf{Ru}_2\mathbf{O}_7$}{Ca2Ru2O7} and \texorpdfstring{$\mathbf{Pr}_2\mathbf{Ru}_2\mathbf{O}_7$}{Pr2Ru2O7}}

\setlength{\tabcolsep}{8pt}
\begin{table}[!htb]
\renewcommand{\arraystretch}{1}
    \caption{Values of $U$ and $J$ under different screening conditions [unscreened (bare) and constrained RPA (cRPA)], bandwidth $W$, and correlation strength $U_{\mathrm{cRPA}}/W$ for Pr$_2$Ru$_2$O$_7$, Ca$_2$Ru$_2$O$_7$, and Cd$_2$Ru$_2$O$_7$. All values except for $U_{\mathrm{cRPA}}/W$ are given in eV.}
    \centering
    \resizebox{0.5\textwidth}{!}{%
    \begin{tabular}{clllllll}
    \hline\hline
     \multicolumn{1}{c}{ } & \multicolumn{1}{c}{$U_{\mathrm{bare}}$} & \multicolumn{1}{c}{$U_{\mathrm{cRPA}}$} & \multicolumn{1}{c}{$J_{\mathrm{bare}}$} & \multicolumn{1}{c}{$J_{\mathrm{cRPA}}$} & \multicolumn{1}{c}{$W$}& \multicolumn{1}{c}{$U_{\mathrm{cRPA}}/W$} \\
        \hline
    Pr$_2$Ru$_2$O$_7$ & 11.492 & 2.393 	&	0.410 	&	0.341 	&	2.000 	&	1.196   \\
    Ca$_2$Ru$_2$O$_7$ & 9.917 &   1.242 	&	0.304 	&	0.244 	&	2.341 	&	0.531   \\
    Cd$_2$Ru$_2$O$_7$ & 9.488  &   1.638 	&	0.289 	&	0.239 	&	2.310 	&	0.709   \\    
    \hline\hline   
        \end{tabular}%
    }
\label{Pr_Ca}
\end{table}

In this section, we analyze the differences between trivalent and divalent pyrochlores, focusing on the critical influence of electronic correlations on the properties of Ca\(_2\)Ru\(_2\)O\(_7\) and R\(_2\)Ru\(_2\)O\(_7\) compounds. Experimental investigations highlight the significant role of electronic correlations in determining the distinct behaviors observed in these materials. Notably, Ca\(_2\)Ru\(_2\)O\(_7\) exhibits metallic resistivity, while Pr\(_2\)Ru\(_2\)O\(_7\) demonstrates insulating characteristics~\cite{PhysRevB.103.L201111,PhysRevB.102.041114}. 
A significant distinction between Ca\(_2\)Ru\(_2\)O\(_7\) and Pr\(_2\)Ru\(_2\)O\(_7\) is the electron occupancy within the Ru-\(t_{2g}\) bands; Ca\(_2\)Ru\(_2\)O\(_7\) possesses 3 electrons, while Pr\(_2\)Ru\(_2\)O\(_7\) contains 4 electrons. 
Experimental evidence indicates that the Ru-\(t_{2g}\) bands in Pr\(_2\)Ru\(_2\)O\(_7\) exhibit stronger correlations relative to the half-filled configuration in Ca\(_2\)Ru\(_2\)O\(_7\), an observation that is atypical for correlated, multi-orbital systems~\cite{10.1146/annurev-conmatphys-020911-125045}. The subsequent sections will delve into the underlying mechanisms driving this counterintuitive behavior.

The derived Hubbard interaction parameters, as detailed in Table~\ref{Pr_Ca}, reveal significant variations in the $U$ parameters between Ca\(_2\)Ru\(_2\)O\(_7\) and Pr\(_2\)Ru\(_2\)O\(_7\). In Pr\(_2\)Ru\(_2\)O\(_7\), the on-site $U$ value is comparable to the bandwidth, whereas in Ca\(_2\)Ru\(_2\)O\(_7\), the $U$ value is considerably smaller. This disparity results in substantial differences in electronic correlations~\cite{Huebsch_2022}.

The factors contributing to the pronounced difference in $U$ values are the spatial extent of the Wannier orbitals and the electronic screening effect. 
In both Ca\(_2\)Ru\(_2\)O\(_7\) and Cd\(_2\)Ru\(_2\)O\(_7\), the nominal valence of the Ru cations is 5+, resulting in a lower energy for the Ru-\(t_{2g}\) orbitals compared to the Pr compound, which has Ru\(^{4+}\) cations, due to the stronger nuclear attractive potential. This causes the energy levels of the Ru-\(t_{2g}\) and O-$2p$ orbitals to be drawn closer together, enhancing the hybridization between these orbitals and leading to more delocalized Wannier functions.
The spatial spread of the Wannier orbitals is reflected in the bare $U$ value listed in Table~\ref{Pr_Ca}: the $U$ values for Ca\(_2\)Ru\(_2\)O\(_7\) and Cd\(_2\)Ru\(_2\)O\(_7\) are indeed much smaller than those for Pr\(_2\)Ru\(_2\)O\(_7\).

Furthermore, the disparity in electronic screening further amplifies the difference in $U$ values beyond the spatial extent of the Wannier orbitals. As indicated in Table~\ref{Pr_Ca}, the disparity in electronic screening effects further amplifies the difference in $U$ values. The \(U_{\mathrm{bare}}\) value for Ca\(_2\)Ru\(_2\)O\(_7\) is larger than that for Cd\(_2\)Ru\(_2\)O\(_7\), but after screening, the \(U_{\mathrm{cRPA}}\) value for Ca\(_2\)Ru\(_2\)O\(_7\) is smaller than that for Cd\(_2\)Ru\(_2\)O\(_7\).
This pronounced screening effect arises because, in Ca\(_2\)Ru\(_2\)O\(_7\), the O-$2p$ bands overlap with the Ru-\(t_{2g}\) manifold and lie very close to the Fermi level [Fig.~\ref{fig5}(a-b)], which substantially contributes to the screening. This effect is due to the reduced energy appearing in the denominator of the polarization function expression, leading to a lower $U$ value for Ca\(_2\)Ru\(_2\)O\(_7\).

\subsection{Pressure dependence for \texorpdfstring{$\mathbf{Ca}_2\mathbf{Ru}_2\mathbf{O}_7$}{Ca2Ru2O7} and \texorpdfstring{$\mathbf{Cd}_2\mathbf{Ru}_2\mathbf{O}_7$}{Cd2Ru2O7}}
\label{A2Ru2O7_sec} 

\begin{figure}[htb!]
\begin{center}
    \includegraphics[width=\columnwidth]{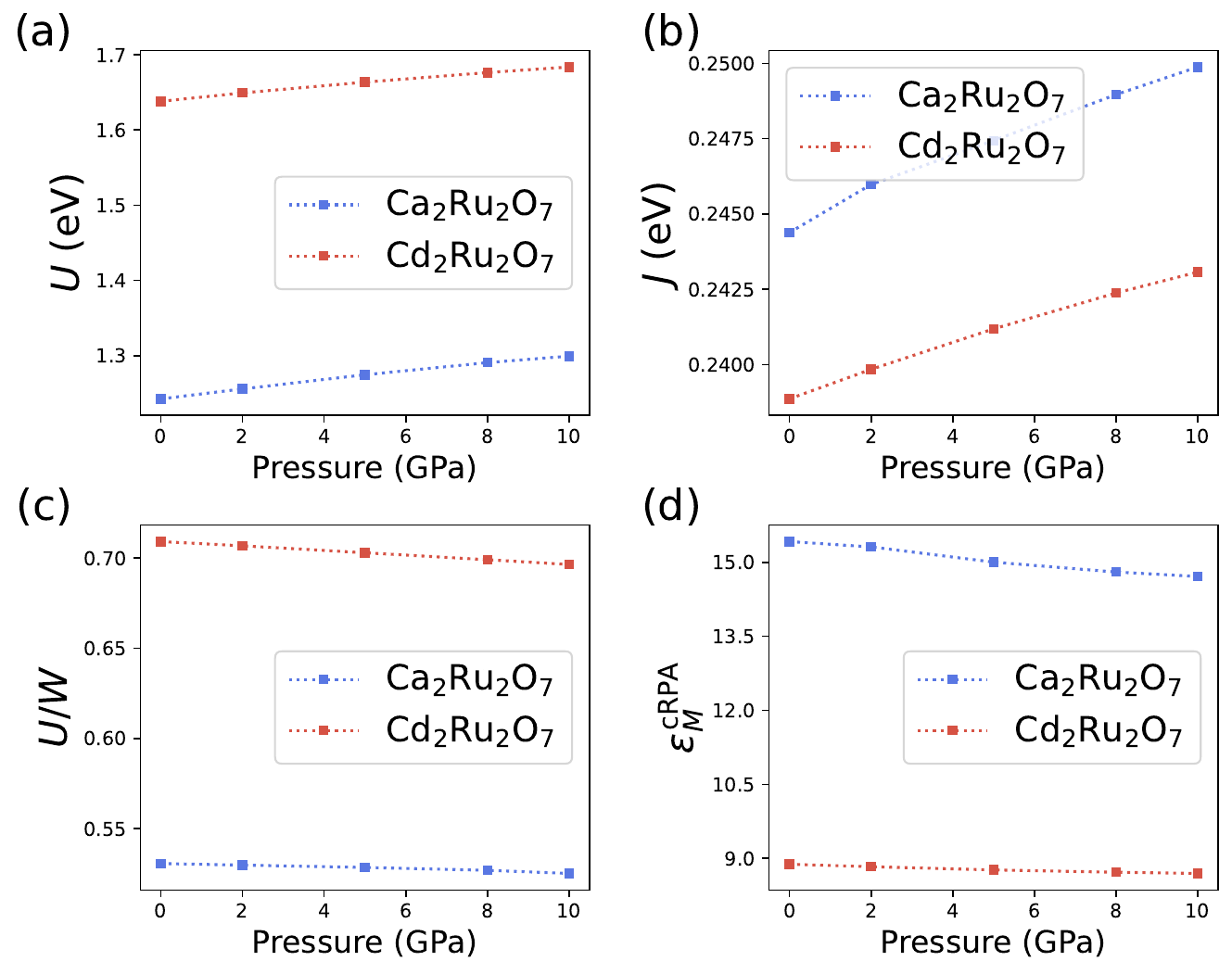}
    \caption{(Color online) The dependence of pressure for Ca$_2$Ru$_2$O$_7$ and Cd$_2$Ru$_2$O$_7$: (a) the   on-site effective Coulomb repulsion \(U\), (b) the on-site effective exchange interaction $J$, (c) the correlation strength $U / W$, which are derived within the cRPA method and (d) cRPA-macroscopic-dielectric constant $\epsilon _{M}^{\mathrm{cRPA}}$.
}
\label{A2Ru2O72}    
\end{center}
\end{figure}

Fig.~\ref{A2Ru2O72} summarizes the results of the cRPA calculations, including the   on-site effective Coulomb repulsion \(U\), the on-site effective exchange interaction $J$, and the correlation strength $U / W$.
Due to the overlap of O-$2p$ bands with the Ru-\(t_{2g}\) manifold in Ca\(_2\)Ru\(_2\)O\(_7\), as mentioned earlier, 
$U$ value of Ca\(_2\)Ru\(_2\)O\(_7\) is always smaller than that of Cd\(_2\)Ru\(_2\)O\(_7\).
The material dependence of $\epsilon _{M}^{\mathrm{cRPA}}$ [Fig.~\ref{A2Ru2O72}(d)] contributes to a corresponding variation in $U$, with the $U$ values for Cd\(_2\)Ru\(_2\)O\(_7\) and Ca\(_2\)Ru\(_2\)O\(_7\) consistently following a similar trend.

We observe that the Hubbard $U$ is increasing with pressure 
 [Fig.~\ref{A2Ru2O72}(a)] and the values of \(U/W\) are almost unchanged as opposed to a naive expectation [Fig.~\ref{A2Ru2O72}(c)]. 
\(W\) usually increases with pressure due to the reduction in interatomic distances, which enhances the overlap between atomic orbitals and broadens the electronic bands. This increase in \(W\) traditionally leads to a decrease in the \(U/W\) ratio, suggesting a transition towards more metallic behavior as pressure is applied. 
However, Fig.~\ref{A2Ru2O72} reveals that while \(W\) increases with pressure as expected, \(U\) also increases at a comparable rate with a decrease in lattice constant [Fig.~\ref{fig4}(a)]. 
One of the reasons for this unusual behavior would be attributed to the presence of O orbital bands near the Fermi level, which significantly alter the electronic screening effects. The proximity of these O orbital bands enhances the screening of the Coulomb interaction, resulting in more sensitive change in $U$ as function of applied pressure. Consequently, the \(U/W\) ratio remains stable over a range of pressures. 

These observed stable trends give a crucial hint in understanding metal-to-insulator transition in Cd$_2$Ru$_2$O$_7$~\cite{PhysRevB.98.075118}. The metallic-like state in Cd$_2$Ru$_2$O$_7$ is sensitive to external perturbations. It can be suppressed by applying approximately 1 GPa of hydrostatic pressure or by replacing 5-10\% of Cd with Ca. This suggests that the electronic states are unstable and can transition from metallic to insulating with slight changes in external conditions. The transition into insulating behavior is unexpected since we usually believe that pressure puts materials into weakly-correlated side. 
However, in this material, pressure has little effect on $U/W$, giving room for other mechanisms to work for the pressure-induced metal-insulator transition.

\section{Discussion and conclusion}
\label{Conlusion}  

We conducted a comprehensive \textit{ab initio} study to calculate all interactions in a multiorbital Hubbard model from first principles, relying solely on atomic positions. Our focus was on the electronic properties of R$_2$Ru$_2$O$_7$, R$_2$Ir$_2$O$_7$, Ca$_2$Ru$_2$O$_7$, and Cd$_2$Ru$_2$O$_7$, where R$^{3+}$ represents a rare-earth ion. This investigation examines how different rare-earth elements influence the electronic behavior, with particular attention to the complex interplay between lattice constants, electronic interactions, and atomic positions that collectively define the material's properties.

Our findings show that \( U \) for R$_2$Ru$_2$O$_7$ and R$_2$Ir$_2$O$_7$ actually increases as the ionic radius of R decreases. Kaneko et al.~\cite{PhysRevB.102.041114} demonstrated that the charge gap of R$_2$Ru$_2$O$_7$ increases and the peak of the Hubbard band shifts upwards as R changes from Pr to Lu. These observations are consistent with our findings on the increasing trend of \( U \). In these compounds, \( U \), rather than the bandwidth \( W \), plays a more decisive role in determining the \( U/W \) ratio (interaction control), challenging the conventional belief that changes in \( W \) are of primary importance (bandwidth control). An intriguing point here is that the control of the interaction is achieved differently from the environment-mediated control discussed for 2D materials~\cite{10.1038/s41699-023-00408-x,10.1021/acs.nanolett.5b05009,PhysRevB.100.161102,PhysRevLett.115.186602}.

Furthermore, the roles of Ru$^{4+}$ and Ir$^{4+}$ ions are critical in defining the electronic properties of these compounds. 
Our results indicate that R$_2$Ru$_2$O$_7$ consistently exhibits a higher \( U/W \) ratio than R$_2$Ir$_2$O$_7$. 
In R$_2$Ir$_2$O$_7$, the \( U/W \) ratio increases as the ionic radius of the rare-earth element decreases. This trend aligns with the observed metallic behavior in compounds with larger R$^{3+}$ ions, where the \( U/W \) ratio is low (\( U/W < 1.1 \)), such as Eu, Sm, and Nd~\cite{PhysRevLett.96.087204, 10.1143/JPSJ.76.043706, 10.1143/JPSJ.80.094701,PhysRevB.93.245120}. In contrast, compounds with smaller ionic radii, like Yb and Ho, tend to be insulating~\cite{PhysRevB.86.014428, 10.1143/JPSJ.80.094701,PhysRevB.93.245120}.
Applying external pressure and chemical pressure can effectively alter the lattice constant, leading to changes in the on-site effective Coulomb repulsion \( U \) due to the change in spatial extension of the orbitals and in screening strength. 
Interestingly, physical pressure and chemical pressure have opposing effects on \( U/W \). When both decrease the lattice constant, physical pressure leads to a decrease in \( U/W \), while chemical pressure causes an increase in \( U/W \). 
This difference arises because the reduction in bond angles due to physical pressure is much smaller than that caused by chemical pressure for a similar change in lattice constant. In the case of applying physical pressure, when the decrease in bandwidth from bond angle reduction is insufficient to offset the increase in bandwidth from lattice constant compression, the increase in \( W \) outpaces that of \( U \), resulting in a decrease in the \( U/W \) ratio. Conversely, in the case of applying chemical pressure, when the change in Ru-O-Ru bond angles is enough to counteract the effects of lattice constant compression, the \( U/W \) ratio will rise.
These approaches allow for the fine-tuning of material properties to achieve a desired level of electron-electron interaction.

For Ca$_2$Ru$_2$O$_7$ and Cd$_2$Ru$_2$O$_7$, the pressure dependence of \( U/W \) is atypical compared to R$_2$Ru$_2$O$_7$ and R$_2$Ir$_2$O$_7$. While \( U \) also changes with pressure, the \( U/W \) ratio remains nearly unchanged, contrary to the common assumption that applying pressure decreases \( U/W \). Our analysis, supported by both calculations and experimental results, suggests that MIT in Cd$_2$Ru$_2$O$_7$ induced by pressure is not primarily driven by changes in \( U/W \), but rather by the intricate interplay of various effects, including spin-orbit coupling, Hund's coupling, and trigonal distortion.
It is of great interest to study further on this complex interplay to reveal a mechanism of the unusual pressure-induced metal-to-insulator transition.

In conclusion, our \textit{ab initio} studies highlight the complexity of interactions in pyrochlore compounds and emphasize the need to consider multiple factors to fully understand the mechanisms driving metal-insulator transitions in these materials. Our work lays the groundwork for further experimental and theoretical investigations into the electronic phases of pyrochlore oxides.

\section*{Acknowledgement}
We acknowledge fruitful discussion with Kentaro Ueda.
This work is supported by the National Natural Science Foundation of China~(Grant No. 12204130), Shenzhen Start-Up Research Funds~(Grant No. HA11409065), Shenzhen Key Laboratory of Advanced Functional CarbonMaterials Research and Comprehensive Application (Grant No. ZDSYS20220527171407017). Y.N. is supported by MEXT as ``Program for Promoting Researches on the Supercomputer Fugaku'' (Grant No. JPMXP1020230411), Grant-in-Aids for Scientific Research (JSPS KAKENHI) (Grant Nos. JP23H04869, JP23H04519, and JP23K03307), and JST (Grant No. JPMJPF2221).

\bibliography{references.bib}
\end{document}